\documentclass[twocolumn,showpacs,preprintnumbers,amsmath,amssymb]{revtex4}
\usepackage{graphicx}
\usepackage{dcolumn}

\begin{document}
\title{Theoretical study of dark resonances in micro-metric thin cells}
\author{H. Failache}
\email{heraclio@fing.edu.uy}
\author{L. Lenci}
\author{A. Lezama}
\affiliation{Instituto de F\'{\i}sica, Facultad de Ingenier\'{\i}a,
Universidad de la Rep\'{u}blica,\\ J. Herrera y Reissig 565, 11200
Montevideo, Uruguay}
\author{D. Bloch}
\author{M. Ducloy}
\affiliation{Laboratoire de Physique des Lasers, UMR 7538 du CNRS,
Institut Galil\'{e}e,\\ Universit\'{e} Paris 13, 99 av. J.-B.
Cl\'{e}ment, 93430 Villetaneuse, France}
\date{\today}
\begin{abstract}
We investigate theoretically dark resonance spectroscopy for a
dilute atomic vapor confined in a thin (micro-metric) cell. We
identify the physical parameters characterizing the spectra and
study their influence. We focus on a Hanle-type situation, with an
optical irradiation under normal incidence and resonant with the
atomic transition. The dark resonance spectrum is predicted to
combine broad wings with a sharp maximum at line-center, that can be
singled out when detecting a derivative of the dark resonance
spectrum. This narrow signal derivative, shown to broaden only
sub-linearly with the cell length, is a signature of the
contribution of atoms slow enough to fly between the cell windows in
a time as long as the characteristic ground state optical pumping
time. We suggest that this dark resonance spectroscopy in
micro-metric thin cells could be a suitable tool for probing the
effective velocity distribution in the thin cell arising from the
atomic desorption processes, and notably to identify the limiting
factors affecting desorption under a grazing incidence.
\end{abstract} \pacs{39.30.+W 32.70.Jz 42.62.Fi 42.65.-K}
\maketitle
\section{\label{sec:level1}Introduction}
Large attention has been devoted during the last years to
coherently prepared atomic media. Coherent Population Trapping
(CPT), Electromagnetically Induced Transparency (EIT),
Electromagnetically Induced Absorption (EIA), sub-recoil laser
cooling, lasing without inversion, strong dispersive media, fast
and slow light are some examples of physical phenomena resulting
from the interaction of light fields with coherently prepared
media.\\
EIT is a nonlinear optical effect typically observed in a
$\Lambda$-type three-level atomic system interacting with two
electromagnetic fields. The significant reduction of the medium
absorption, occurring near the Raman resonance condition, can be
understood as the consequence of the existence of a coherent
superposition of ground levels (a dark state) which is not coupled
to the excited state by the fields. Atoms falling in this dark state
stay trapped as the electromagnetic field is unable to excite them.
Transparency is then induced as a consequence of the optical pumping
of the atomic system into this dark state. In consequence, EIT
resonances are commonly designated as dark resonances as we will do
in the following. The width of the dark resonance when the Raman
detuning is varied is essentially determined by the coherence loss
rate of the lower (ground) states. In closed $\Lambda $ systems, the
optical pumping rate to the dark state defines a lower limit for the
resonance width \cite{Arimondo:1996b}. However a quite different
behavior is observed for open systems, where the lower limit of the
resonance width (in the absence of other dephasing processes)
appears to be only determined by the interaction time of the atoms
with the laser radiation \cite{Renzoni:1998}. For an atomic vapor
inside a macroscopic cell the interaction time is essentially
determined by the time of flight of atoms through the laser beam. A
buffer gas, resulting in a diffusive atomic motion, is frequently
used in order to largely increase such a time. Another approach
consists in avoiding the atom coherence loss at the walls of the
container by using a paraffin coating. In practice, other factors
such as collisions or magnetic field inhomogeneities can result in
additional ground state decoherence. Dark resonance spectra in
macroscopic cells can be extremely narrow, spectral widths smaller
than 50 Hz have been measured in buffered cm-long cells
\cite{Brandt:1997,Merimaa:2003}.\\
Considerable attention is presently directed to the use of
miniaturized atomic cells as practical metrological standards. Dark
resonances in miniaturized cells have been used as compact atomic
frequency references \cite{Lutwak:2003,Knappe:2004} and compact
magnetometers \cite{Schwindt:2004}. However, in order to obtain
narrow resonances with such miniaturized cells, large buffer gas
pressures must be used to provide a small enough atomic diffusion
lengths. The use of miniaturized buffer gas cells has the drawback
that the dark resonance linewidth grows very fast when the cell
characteristic dimension $L$ is reduced (width$\propto L^{-2}$
\cite{Vanier:1989}). The spectroscopic properties of small size
cells can be considerably different than those of larger ones if the
cell dimensions are small compared to the mean free path of the
atoms in the vapor.  An example of this is given by the linear
absorption spectrum of an atomic gas with 1D confinement between two
close parallel windows obtained in a thin cell. In this case the
atomic sample becomes anisotropic since atoms flying parallel to the
windows interact with light during a much larger time than those
flying perpendicular to the windows. Such an anisotropy is
responsible for the observation of sub-Doppler features in the
spectrum \cite{Briaudeau:1996,Briaudeau:1999}.\\
Dilute atomic vapor cells, in which the atomic trajectories are from
wall-to-wall, can be classified in three different classes depending
on the interaction time of an atom flying at the mean thermal
velocity $\bar{v}$. ``Long'' cells are those where the average time
of flight exceeds the longest characteristic time of the atomic
evolution generally given by an optical pumping time $\tau_p$, i.e.
$L \gg \bar{v}\tau_p$. In such cells, an atom departing from one
wall reaches a steady state regime before hitting the opposite cell
wall. Intermediate cells, designated as ``micro-metric'' cells
\cite{Briaudeau:1998} are those for which $\bar{v}\tau_p \gtrsim
L\gtrsim \frac{\bar{v}}\Gamma $ ($\Gamma^{-1}$: excited state
lifetime). In these cells the atom-light interaction time is long
compared to the excited state decay which means that several
absorption-emission cycles can take place. However, the optical
pumping efficiency is strongly dependent on the atomic velocity and
only the slowest atoms may approach a steady state. Finally,
nanometric cells are those for which $L\lesssim \frac{\bar{v}}\Gamma
$ which means that the optical relaxation can be neglected during
the atomic flight, the regime is a coherent build-up of the atomic
excitation for those cells whose length is typically shorter than
the optical wavelength.\\
Micro-metric thin cells were used in optical pumping experiments
with a single irradiating beam at normal incidence. The strong
dependence of the interaction time with the atomic velocity results
in the observation of Doppler free absorption spectral lines
\cite{Briaudeau:1996}. Nano-metric cells have been used to observe
the enhancement of the coherent transient atomic response due to
Dicke narrowing, and to study long-range atom-wall interaction
\cite{Dutier:2003b,Fichet:2007}. Recently, a merging between the
domain of EIT in miniaturized cells and spectroscopy in anisotropic
thin cells, has led to the observation of dark resonances in
micro-metric thin cells \cite{Fukuda:2003,Failache:2004,Knappe:2004}
and even in nano-metric thin cells
\cite{Varzhapetyan:2005,Sargsyan:2006}. It is the purpose of the
present paper to describe theoretically the main spectroscopic
features of dark resonances in micro-metric thin
cells\cite{Izmailov:1996,Izmailov:1998,Izmailov:1999,Petrosyan:2000},
with an emphasis on the contribution of slow atoms.\\

\section{\label{sec:level2}Theoretical study}
\subsection{\label{sbsec:level3}Atomic model}

\begin{figure}
\includegraphics[width=8.6cm]{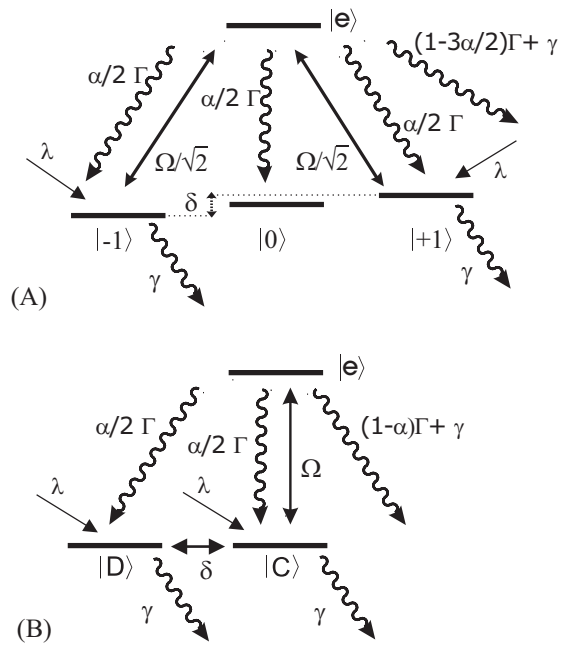}
\caption{\label{fig:levels} A) Atomic model, transition $F=1
\rightarrow F'=0$ B) Elementary 3-level atomic model in the base
$|C\rangle$, $|D\rangle$.}
\end{figure}

We have chosen to study the properties of dark resonances in
micro-metric cells through the analysis of a scheme applicable to
the experimental situation of an atomic transition with a lower
level with total angular momentum $F=1$ and an excited state with
angular momentum $F'=0$ interacting with a single linearly polarized
optical field propagating along the quantization axis. With the
incident field expanded into the components of the two eigen
circular polarization components, this system constitutes a
realization of a $\Lambda$ system (see Fig.~\ref{fig:levels}.A). The
Raman detuning can be easily tuned with the help of a magnetic field
collinear with the light beam. Similar configurations, usually
designated as Hanle/CPT schemes, have been used in many EIT
experiments
\cite{Arimondo:1996b,Renzoni:1998,Renzoni:1999,Dancheva:2000,Valente:2002}.
The system is open since the atoms can decay from the excited state
into the $|F=1,m_{F}=0\rangle$ or into external states. The choice
of this configuration is motivated by its large symmetry (equal
atom-field coupling and decay rates on the two arms of the $\Lambda$
system) which allows to simplify the discussion due to a reduction
in the number of independent parameters. Although our study is based
on a specific case, our conclusions can be extended to the general
case of an asymmetric $\Lambda$ system.\\
The dynamics of the atomic system can be conveniently analyzed in
the basis presented in Fig.~\ref{fig:levels}.B where the lower
states are the dark
$|D\rangle\equiv\frac{1}{\sqrt{2}}(|1\rangle-|-1\rangle)$ and
coupled $|C\rangle\equiv\frac{1}{\sqrt{2}}(|1\rangle+|-1\rangle)$
states \cite{Aspect:1988,Aspect:1989}.
The optical fields couple the $|e\rangle $ and $|C\rangle $ sates
with the Rabi frequency $\Omega$. For nonzero Raman detuning
$\delta$ (corresponding to twice the Zeeman splitting introduced by
the magnetic field) the $|D\rangle $ and $|C\rangle $ states are
coupled. Population losses to external levels (including the
$|F=1,m_{F}=0\rangle$ level in the precise case of a $|F=1\rangle
\leftrightarrow |F=0\rangle$ transition), are described by a rate
(1-$\alpha )\Gamma $, with $\Gamma$ the optical width of the
transition and $\alpha$ a branching ratio. The case of a closed
system corresponds to $\alpha = 1$. Aside from a possible population
loss -controlled by the parameter $(1-\alpha)$- a phenomenological
decay rate $\gamma $ is introduced in the model to mimic the
experimental conditions (finite beam diameter, field
inhomogeneities, etc.) that set a lower limit to the lower level
decoherence rate \cite{Arimondo:1996b}. The arrival of fresh atoms
to the system (essentially those leaving from the cell walls) is
represented by constant and equal repumping terms ($\lambda$) for
the lower levels. If the atomic density matrix is nomalized to
unity, we have $2\lambda=\gamma$. The Bloch equations describing the
atom-field interaction in the $|e\rangle $, $|C\rangle $ basis (in
the rotating wave approximation) are \cite{Renzoni:1999},
\begin{subequations}
\label{eq_1}
\begin{eqnarray}
\dot{\sigma}_{DD} &=&\alpha \frac{\Gamma}{2} \sigma
_{ee}+\delta\;Im\;\sigma _{DC}-\gamma \sigma _{DD}+\lambda
\end{eqnarray}
\begin{eqnarray}
\dot{\sigma}_{CC} &=&\alpha \frac{\Gamma}{2} \sigma _{ee}-\delta
\;Im\;\sigma _{DC}+2\Omega Im\;\sigma _{eC}- \nonumber\\ &&\gamma
\sigma _{CC}+\lambda
\end{eqnarray}
\begin{eqnarray}
\dot{\sigma}_{ee}=-\Gamma \sigma _{ee}-2\Omega Im\;\sigma
_{eC}-\gamma \sigma _{ee}
\end{eqnarray}
\begin{eqnarray}
\dot{\sigma}_{eD} &=&-(\frac \Gamma
2-i\Delta)\sigma_{eD}-i\frac{\delta}{2} \sigma _{eC}-i\Omega \sigma
_{CD}- \nonumber\\ &&\gamma \sigma _{eD}
\end{eqnarray}
\begin{eqnarray}
\dot{\sigma}_{eC} &=&-(\frac \Gamma
2-i\Delta)\sigma_{eC}-i\frac{\delta}{2} \sigma _{eD}-i\Omega (\sigma
_{CC}-\sigma _{ee})- \nonumber\\ &&\gamma \sigma _{eC}
\end{eqnarray}
\begin{eqnarray}
\dot{\sigma}_{DC}&=&i\frac{\delta}{2}(\sigma_{CC}-\sigma_{DD})+i\Omega
\sigma _{De}-\gamma \sigma _{DC} \label{eq15}
\end{eqnarray}
\end{subequations}
where $\Delta =\delta \omega - k v_z$ is the optical detuning, $v_z$
being the component of the atomic velocity perpendicular to the cell
window, $k$ the amplitude of the wave vector characterizing the
incident irradiation and $\delta\omega$ the detuning of the laser
with respect to the optical tansition for an atom at rest. The
steady state solution of Bloch equations (Eqs.~\ref {eq_1}) is not
useful in the present case because the atoms contributing to the
coherence resonance are essentially in transient regime of
interaction. In consequence, the Bloch equations should be
integrated to obtain the atomic density matrix $\sigma (t)$ as a
function of the interaction time which is the time elapsed after the
atom has left a cell window. The light intensity absorbed by the
atomic vapor in a thin cell of thickness $L$ is given by
\cite{Zambon:1997,Briaudeau:1999}:
\begin{eqnarray}
\Delta I=\kappa \Omega \int_{-\infty}^\infty
\int_{0}^{L}W(v_z)Im\sigma_{eC}(z,v_z)\;dz\;dv_z \label{eq_5}
\end{eqnarray}
where $z$ is the distance perpendicular to the cell window (along
laser propagation) and $\kappa$ is a coefficient proportional to the
atomic density (an optically thin atomic medium will be assumed
along this article). In Eq.~\ref{eq_5} the interaction time does not
appear explicitly; $z$ is related to $t$ through $z=v_z t$, for the
atoms leaving the cell surface at $z=0$ and $z-L=v_z t$ for atoms
leaving the cell surface at $z=L$. $W(v_z)$ is the velocity
distribution function assumed to be the Maxwell-Boltzmann (M-B)
distribution. Note that when the laser is taken at resonance with
the optical transition ($\delta \omega=0$), as will be assumed along
this article except in section \ref{sbsec:level7}, the contribution
of positive and negative velocities in Eq.~\ref{eq_5} are equal.
Such an assumption ensures that the light is resonant with the atoms
with small velocity component $v_z$ in the direction perpendicular
to the cell windows. This assumption does not represent a strong
experimental restriction since it is enough for the laser detuning
to satisfy $\delta\omega < \Gamma$, as will be discussed in
section~\ref{sbsec:level7}. Multiple reflections of the laser beam
at the cell windows modify the atomic absorption due to a
Fabry-Perot effect. Such an effect is small if the cell length $L$
is longer than the laser wavelength and will be neglected here
\cite{Dutier:2003a}. Boundary effects, such as the finite light beam
diameter, are assumed to be taken into account by the relaxation
rate $\gamma $.\\
\subsection{\label{sbsec:level4}Characteristic parameters and lineshape predictions}
\begin{figure}[!]
\includegraphics[width=8.6cm]{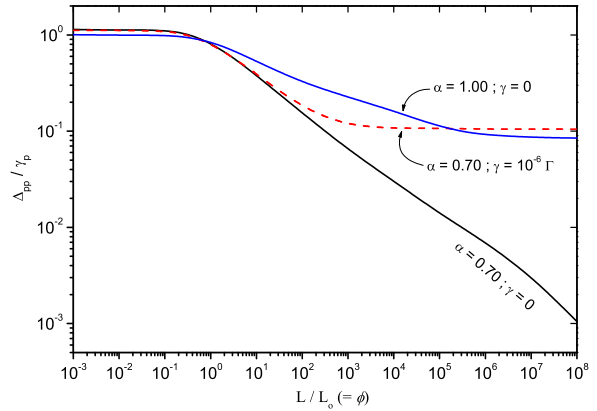}
\caption{\label{fig:width}(Color online) Peak-to-peak width of the
derivative of the dark resonance spectra as a function of the cell
thickness $L$($\Omega=0.01 \Gamma$). Solid lines: $\lambda=\gamma
=0$ for an open system with $\alpha=0.7$ and a closed system
($\alpha\equiv1$). Dashed line: $\lambda=\gamma =10^{-6}\Gamma$
($\alpha=0.7$).}
\end{figure}

\begin{figure}[!]
\includegraphics[width=8.6cm]{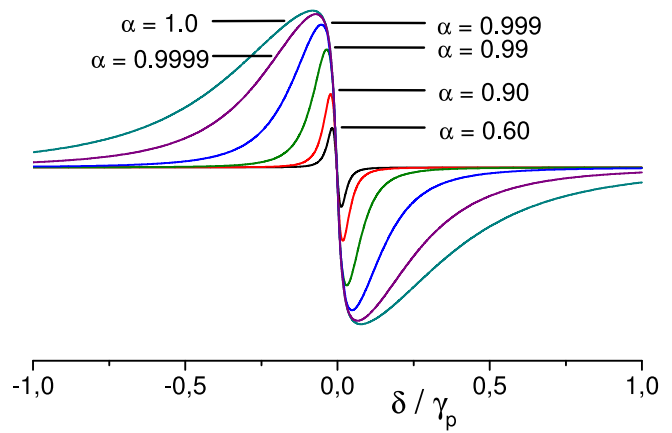}
\caption{\label{fig:narr}(Color online) Influence of the population
losses (determined by the parameter $\alpha$) on the spectral
narrowing for $\phi=10^4$ and $\lambda=\gamma=0$.}
\end{figure}

\begin{figure}[!]
\includegraphics[width=8.6cm]{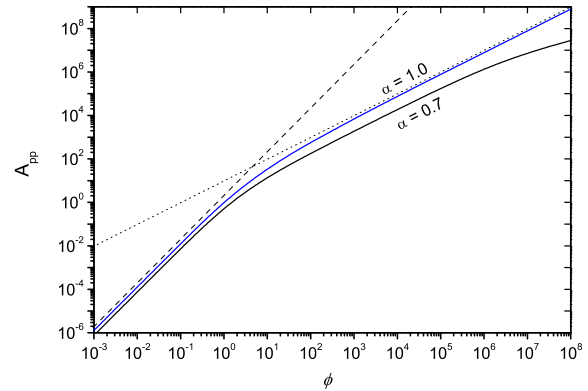}
\caption{\label{fig:amp}(Color online) Theoretical peak-to-peak
amplitude $A_{pp}$ of the derivative of the dark resonance spectra
calculated for different values of the parameter $\phi$
($\Omega=0.01 \Gamma$, $\alpha=0.7, 1.0$, $\lambda=\gamma =0$). The
values for $A_{pp}$ were taken relative to its value at $\phi = 1;
\alpha = 1$. The dashed line illustrates an asymptotic evolution
proportional to $\phi^2$ while the dotted line corresponds to a
linear dependence on $\phi$.}
\end{figure}

We consider here the simple situation $\gamma = \lambda = 0$ (i.e.
the loss of population and arrival of fresh atoms are neglected on
the time scale involved in the atom-light interaction; the role of
$\gamma$ is considered in ~\ref{sbsec:level6}). From the analytic
expression of the eigenvalues of the matrix associated to the first
order differential equations system (Eq.\ref{eq_1}) it is possible
to identify dimensionless parameters characterizing the dark
resonance spectra in a thin cell. The corresponding discussion
together with the numerical verification is deferred to the
Appendix. We find that the atomic response, as a function of the
Raman detuning, depends on the cell length $L$ and the light
intensity through the dimensionless characteristic parameter
\begin{subequations}
\label{eq_phi}
\begin{eqnarray}
\phi =\frac{ \Omega ^2kL}{\Gamma ^2}
\end{eqnarray}
with
\begin{eqnarray}
\frac{\Omega ^2}\Gamma = \gamma_p
\end{eqnarray}
\end{subequations}
where $\gamma_p$ is the optical pumping rate from the $|C\rangle $
state. Note that $\gamma_p L$ represents the maximal velocity
allowing an atom to hit the opposite wall after an efficient
pumping, while $\frac{\Gamma}{k}$ is the velocity width for optical
processes of velocity selection. The parameter $\phi$, already
noticed as a characteristic one in single beam optical pumping
experiments \cite{Briaudeau:1996}, appears to be the ratio between
the specific kind of velocity "selection", as allowed by the optical
pumping, over the classical optical velocity selection. The atomic
response depends also on the branching ratio $\alpha$ and the
dimensionless Raman detuning $\frac{\delta}{\gamma_p}$.\\
\begin{figure}[!]
\includegraphics[width=8.6cm]{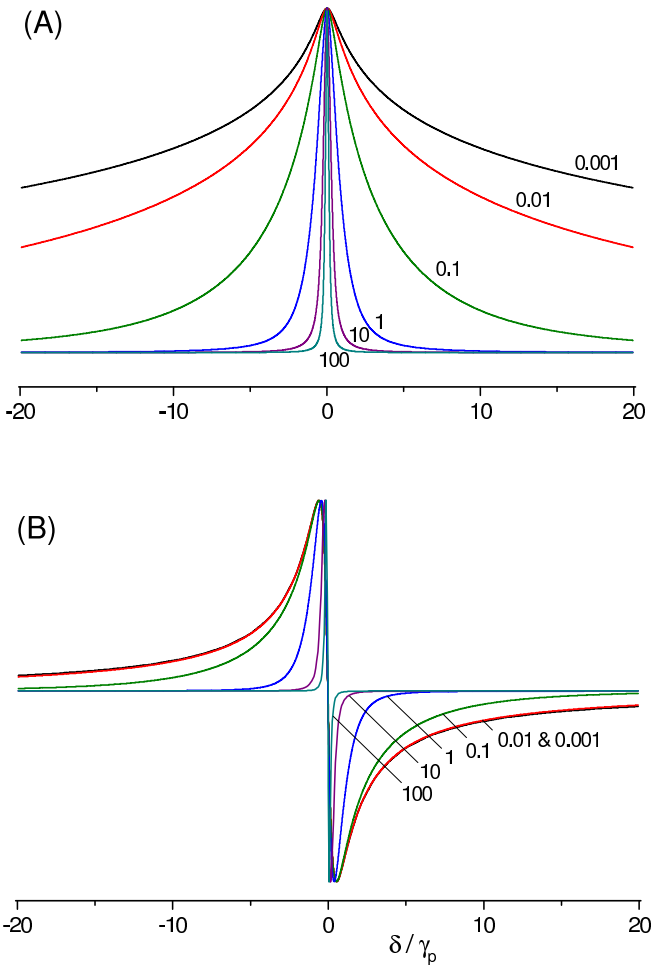}
\caption{\label{fig:theo}(Color online) A) Dark resonance spectra
for different values of the parameter $\phi$ ($\Omega=0.01 \Gamma$,
$\alpha=0.7$, $\lambda=\gamma =0$). For the purpose of line-shape
comparison, the amplitude of spectra are normalized. (The vertical
axis corresponds to transmission). B) Derivative with respect to the
Raman detuning of the spectra shown in A). The corresponding
spectrum amplitudes $A$, relative to the one for $\phi=1$, are: A)
$A_{0.001}=1.0\times10^{-5}$, $A_{0.01}=7.5\times10^{-4}$,
$A_{0.1}=3.8\times10^{-2}$, $A_1=1$, $A_{10}=9.9$, $A_{100}=56.4$ ;
B) $A_{pp\:0.001}=1.4\times10^{-6}$,
$A_{pp\:0.01}=1.4\times10^{-4}$, $A_{pp\:0.1}=1.4\times10^{-2}$,
$A_{pp\:1}=1$, $A_{pp\:10}=28.6$, $A_{pp\:100}=367.$}
\end{figure}
\begin{figure}[!]
\includegraphics[width=8.6cm]{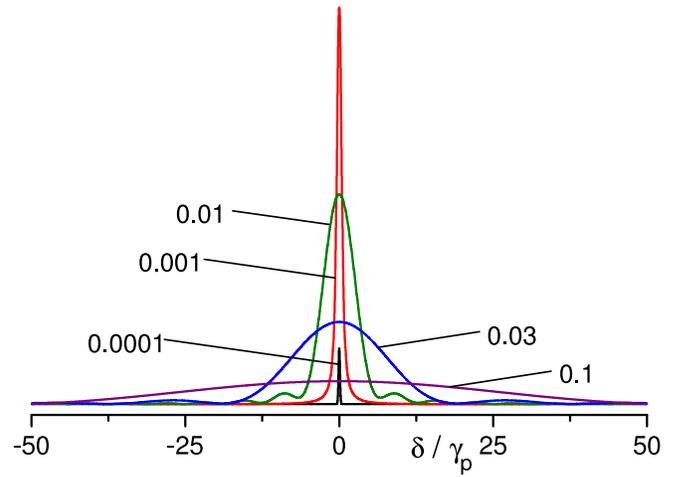}
\caption{\label{fig:vel_cont}(Color online) Contribution of
different velocities $v_z$ to a dark resonance spectrum. The
vertical axis corresponds to transmission. The choice of parameters
$\phi=0.01$, $\alpha=0.7$, $\lambda=\gamma =0$ corresponds to a
relatively broad spectrum. The figures on each curve correspond to
the velocity in $\frac{\Gamma}{k}$ unit.}
\end{figure}
As expected it is the optical pumping rate $\gamma_p$ and not the
strength of the optical coupling $\Omega$, that determines the
relevant time-scale of the atomic evolution. All over the paper we
will assume $\Omega\ll \Gamma $. This is because the interesting
effects for the dark resonance should not be altered by the
saturation and broadening of the optical transition. Hence, the
transient effects on the optical coherence, at the heart of
nanometric cell behavior, can be neglected. This also implies that
large $\phi$ values are associated to relatively
long cell lengths.\\
Fig.~\ref{fig:theo} shows the dark resonance spectra for different
values of the characteristic parameter $\phi$, the other parameters
being kept constant. In Fig.~\ref{fig:theo} and generally all along
this work, the dark resonance spectrum is isolated from the spectrum
obtained from Eq.\ref{eq_5} subtracting the linear absorption and
the single beam optical pumping spectra
\cite{Briaudeau:1996} broad background calculated separately.\\
The shape of the dark resonance spectra for small values of $\phi$
is characterized by slowly decreasing wings and a sharp peak at
$\delta =0$. Although the width and shape of the spectra
considerably evolve with the parameter $\phi$, one notes a
persistent singularity at line center. Such a sharp central feature
is better seen (Fig.~\ref{fig:theo}.B) when considering the
derivative of the spectrum with respect to the Raman detuning, as
can be practically obtained through a demodulation of a modulated
Zeeman structure. In the following, we mostly concentrate on the
analysis of this derivative of the spectrum.\\
Slow atoms, having long time of interaction, contribute to the
coherent signal through narrow resonances while faster atoms
essentially contribute to the broad wings of the coherent signal
(see Fig.~\ref{fig:vel_cont}). Although slow atoms are rare, their
contribution  at line-center is of comparable importance to that of
the more abundant fast atoms. Moreover, it is the narrower
contribution of slow atoms that dominates the spectrum derivative.
The detailed behavior of the resonances can be notably illustrated
through the evolution of the peak-to-peak width $\Delta_{pp}$ of the
derivative of the dark resonance spectra as a function of the
dimensionless parameter $\phi$ (Fig.~\ref{fig:width}). In a more
practical manner, Fig.~\ref{fig:width} shows the influence of the
cell thickness $L$. Indeed, $\phi$ is proportional to $L$, so that
$\phi$ can be seen as a length ratio $\frac{L}{L_o}$ when a given
Rabi frequency $\Omega$ is assumed ($L_o$ is a characteristic length
defined as $L_o\equiv\frac{\Gamma}{k \gamma_p}$, it represents the
length above which the optically selected atoms with $|v_z| <
\frac{\Gamma}{k}$ approach the steady-state relatively to the Raman
coherence). Two different regimes can be distinguished according to
$L \lesssim L_o$ or $L \gtrsim L_o$ (i.e. $\phi \lesssim 1$ or $\phi
\gtrsim 1$). In addition, differences between closed and open
systems appear when $L \geq L_o$ (i.e. $\phi \geq 1$).\\
For the thinnest cells it can be seen that the spectral width reach
a maximum $\Delta_{pp} \thickapprox \gamma_p$. Such a maximal width
is imposed by the interaction time, of the order of one pumping
cycle $\gamma_p^{-1}$, required to reach a significant dark state
preparation (in this regime ($\phi \lesssim 1$) every atom
contributing to the spectrum is optically resonant and experiences
the same $\gamma_p$). The fact that the width $\Delta_{pp}$ remains
finite even for extremely thin cells, critically relies on the
existence of atoms with arbitrarily slow longitudinal velocity. The
existence of such slow atoms (as slow as the optically selected
velocity $\frac{\Gamma}{k}$ times the arbitrary small ratio
$\frac{L}{L_o}$), granted by the assumed M-B velocity distribution
which has an almost flat dependence on $v_z$ around $v_z=0$, is
actually questionable and discussed in ~\ref{sec:level8}.\\
Increasing the cell thickness, the spectral width decreases for $L
\gtrsim L_o$ (i.e. $\phi \gtrsim 1$). Various behaviors thus appears
in Fig.~\ref{fig:width} according to the atomic relaxation
processes, including the closed or opened nature of the system. As
was shown by Renzoni and coworkers \cite{Renzoni:1998}, the
population loss in open systems results in a continuous decrease of
the dark resonance width with the interaction time. This narrowing
is due to the fact that population losses increase with detuning. In
consequence, it reduces the spectral wings with respect to the
central feature (see Fig.~\ref{fig:narr}).\\
Closed atomic systems should be analyzed differently as the dark
resonance width in a closed system is determined by the pumping rate
$\gamma_p$ \cite{Arimondo:1996b}. The spectral width $\Delta_{pp}$
experiences slight variations (of at most one order of magnitude) as
a function of $L$ between the upper and lower limits, both related
to the pumping rate $\gamma_p$. The slight narrowing of the spectra
observed for $L \gtrsim L_o$($\phi \gtrsim 1$) can be understood if
one includes the contribution of atoms Doppler-detuned from
resonance ($kv_z \gtrsim \Gamma$). In this regime, nonresonant atoms
may have enough time to be optically pumped into the dark state.
However, the optical pumping rate for these Doppler-detuned atoms is
smaller than $\gamma_p$ resulting in a narrower contribution to the
dark resonance spectra as emphasized in the spectrum derivative.
Although the contribution of a nonresonant atom is smaller compared
to that of a resonant one, nonresonant atoms are abundant. The total
number of atoms effectively contributing to the dark resonance
spectrum increases with $L$ up to a limit reached when all atomic
velocities are involved. In our calculations (carried for a Doppler
width $\Delta_D=50\Gamma$) this limit is reached for $L \approx 10^6
L_o$.\\
\begin{figure}[!]
\includegraphics[width=8.6cm]{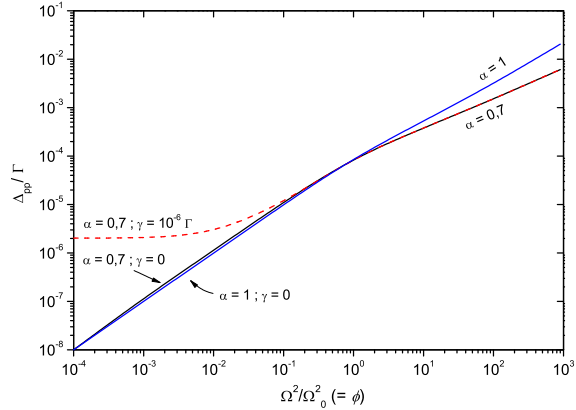}
\caption{\label{fig:widthintens}(Color online) Peak-to-peak width of
the derivative of the dark resonance spectra as a function of the
laser intensity $\Omega^2$($kL=10000$). Solid lines: $\lambda=\gamma
=0$ for an open system with $\alpha=0.7$ and a closed system
($\alpha\equiv1$). Dashed line: $\lambda=\gamma =10^{-6}\Gamma$
($\alpha=0.7$).}
\end{figure}
In the numerical calculation we find that the narrowing of the dark
resonance spectra for an open atomic system follows a $L^{-S}$
($L>L_o$) dependence on the cell thickness with $S \cong \frac{1}{3}
$. Such a dependence suffers only slight variations with the
branching ratio $\alpha$ (provided that $\alpha \lesssim 0.9$).
Hence, the dark resonance spectral width in thin cells is less
sensitive to confinement than an ideal spherical cell with diameter
$L$ for which the mean interaction time is proportional to $L$ and
leads to the well known $\Delta_{pp} \propto L^{-1}$ dependence.
Moreover, in the case of a spherical cell filled with a buffer gas,
as a consequence of the diffusive motion of atoms, the interaction
time is proportional to $L^2$ and the width of the dark resonance
spectrum presents an $L^{-2}$ dependence \cite{Vanier:1989}.\\
Alternately, the dependence on the dimensionless parameter $\phi$
can be discussed as a dependence on the irradiation laser intensity,
as illustrated in Fig.~\ref{fig:widthintens}. The apparent
differences between Fig.~\ref{fig:width} and
Fig.~\ref{fig:widthintens} is related to the definition of the
reduced coordinates for the vertical axis, as
$\frac{\Delta_{pp}}{\gamma_p}$ now depends on $\Omega^2$. The two
regimes $\phi < 1$ and $\phi > 1$ now appear as two regimes for the
laser intensity $\Omega^2 \lesssim \Omega_o^2$ or $\Omega^2 \gtrsim
\Omega_o^2$, where we introduce $\Omega_o^2= \frac{\Gamma^2}{k L}$
($\Omega^2 = \phi \Omega_o^2$) ($\Omega_o$ represents, for a given
cell thickness $L$, the Rabi frequency for which the whole optically
selected zero-velocity class contributes to the coherence
resonance). If $\Omega^2 \lesssim \Omega_o^2$ ($\phi \lesssim 1$)
the atoms are essentially in a transient regime of interaction and
the previously mentioned narrowing process in open systems is not
effective \cite{Renzoni:1998}. In consequence, no essential
difference is found between open or closed atomic systems, and the
spectral width is determined by the pumping rate $\gamma_p$. In this
case every atom contributing to the dark resonance spectrum is
resonant with the optical field and affected by the same optical
pumping rate $\gamma_p$. The spectral width $\Delta_{pp}$ is then
proportional to $\gamma_p$ (see Fig.~\ref{fig:widthintens}). If
$\Omega^2 \gtrsim \Omega_o^2$, for an open atomic system, the
narrowing process already mentioned operates for every atom at
resonance resulting in a slower increase of the spectral width with
the laser intensity ($\Delta_{pp} = (\Omega^2)^{2S}$, $S \cong
\frac{1}{3}$ for $\alpha \lesssim 0.9$). In the case of a closed
atomic system, no narrowing process operates and the spectral width
is determined by the pumping rate. The contribution of atoms detuned
from the optical resonance (that have a smaller effective pumping
rate $\gamma_p$) slows down the evolution of $\Delta_{pp}$ with the
laser intensity. When all atomic velocities contribute to the
coherence resonance (at $\Omega^2 \cong 10^6\;\Omega_o^2$) the
spectral width grows again proportionally to $\gamma_p$.\\
To understand to which extent the predicted narrow lineshapes can be
effectively observed, it is needed to know also the evolution of the
signal peak-to-peak amplitude $A_{pp}$ with respect to the
dimensionless parameter $\phi$ (Fig.~\ref{fig:amp}). Again, one can
discriminate two regimes $\phi \lesssim 1$ and $\phi \gtrsim 1$ and
the respective evolution. For the first case the amplitude grows as
$\phi^2$; for $\phi \gtrsim 1$, it grows proportionally to $\phi$. A
deviation from this behavior is observed for an open system ($\alpha
= 0.7$) around $\phi \gtrsim 10^6$, when the full Doppler velocity
distribution contributes to the dark resonance spectrum. The
peak-to-peak amplitude $A_{pp}$ has a larger amplitude for a closed
atomic system. This shows that the spectral narrowing of the dark
resonance spectrum in an open system, that tends to increase the
amplitude of the derivative, does not compensate for the population
losses responsible for the narrowing mechanism \cite{Renzoni:1998}.\\
\subsection{\label{sbsec:level5} Velocity selection.}
\begin{figure}[!]
\includegraphics[width=8.6cm]{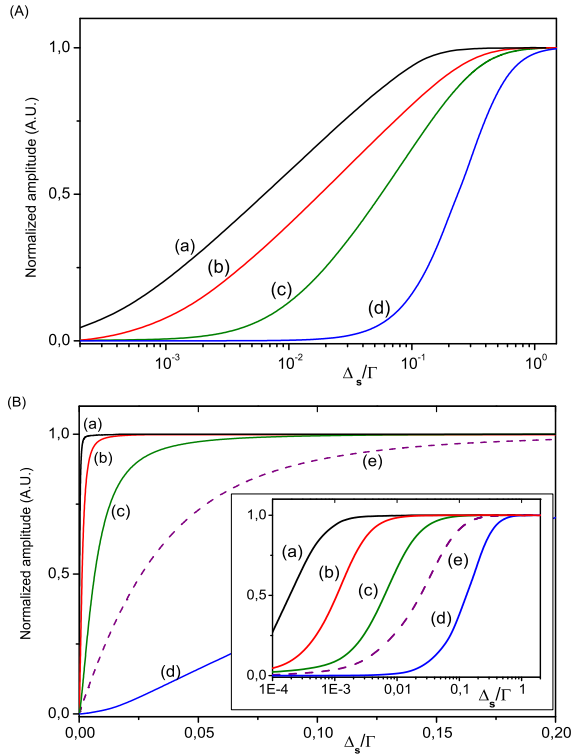}
\caption{\label{fig:velosel}(Color online) A) Partial velocity
contribution to the direct (not derived) dark resonance spectra in a
thin cell corresponding for (a) $\phi=0.001$, $\gamma=0$ (b)
$\phi=0.01$, $\gamma=0$, (c) $\phi=0.1$, $\gamma=0$ and (d)
$\phi=1$, $\gamma=0$, . B) Partial velocity contribution to the
derivative of the dark resonance spectra in a thin cell for the
curves (a-d) shown in A) and (e) $\phi=0.01$, $\gamma
L=0.1\frac{\Gamma}{k}, \frac{\gamma}{\gamma_p}=10$. The data was
normalized to the maximum amplitude.}
\end{figure}

The velocity selection performed by the coherent signal can be
conveniently analyzed by considering the contribution of different
velocity intervals to the dark resonance spectrum as given by the
integral of Eq.~\ref{eq_5}. Following \cite{Briaudeau:1999}, the
contribution of the various velocities to the coherent signal was
analyzed by considering a partial integral for $|v_z| <
\frac{\Delta_s}{k}$. Here again, two different situations should be
considered depending on $\phi<1$ or $\phi>1$.\\
When $\phi<1$ every atom contributing to the dark resonance spectrum
is resonant with the optical transition. Fig.~\ref{fig:velosel}.A
shows the evolution of the amplitude of the dark resonance spectrum
as a function of $\Delta_s$. It can be seen that the whole velocity
interval $\frac{\Gamma}{k}$ ($\frac{\Delta_s}{\Gamma} = 1$) is
essentially responsible for the observed signal. Almost no
contribution to the amplitude of the spectra is due to velocities
larger than $\frac{\Gamma}{k}$. However, if the derivative of the
dark resonance spectrum is considered, the velocities contributing
to this signal are well estimated by the smaller quantity
$\phi\frac{\Gamma}{k}$ as can be seen in Fig.~\ref{fig:velosel}.B.
Hence, almost no contribution to this spectrum arise from atoms out
of the interval $\phi\frac{\Gamma}{k}$. The fact that the derivative
of the spectrum is dominated (for $\phi < 1$) by the contribution of
slow atoms with velocities $|v_z| < \frac{\Gamma}{k}$
(Fig.~\ref{fig:vel_cont}) is a signature of a "logarithmic
singularity" \cite{Schuurmans:1976,Vartanyan:1995,Briaudeau:1999}
occurring when the contribution of large velocities to the direct
spectrum decays
as slowly as $\frac 1{v_z}$.\\
For $\phi < 1$ we conclude that while, strictly speaking, every atom
resonant with the optical transition (i.e. atoms in the velocity
interval $\frac{\Gamma}{k}$) participates in the signal, the idea of
a velocity selection remains a meaningful concept and the velocity
range contributing to the spectra is well estimated by
$\phi\frac{\Gamma}{k}$ as far as the spectrum derivative is
concerned.\\
For $\phi>1$, a fraction of the atoms that have enough time to
contribute to the dark resonance spectrum before hitting the cell
walls are actually fast enough to be Doppler-detuned from the
frequency of the optical field. Such atoms effectively contribute to
the dark resonance spectrum, as was already discussed in the
preceding section, but the velocity interval participating in the
spectra is largely over-estimated by $\phi\frac{\Gamma}{k}$. The
full Doppler velocity distribution can however contribute to the
spectrum for large enough values of $\phi$ (e.g. $\phi \cong
10^6$).\\
\begin{figure*}[!]\centering
\includegraphics[width=16cm]{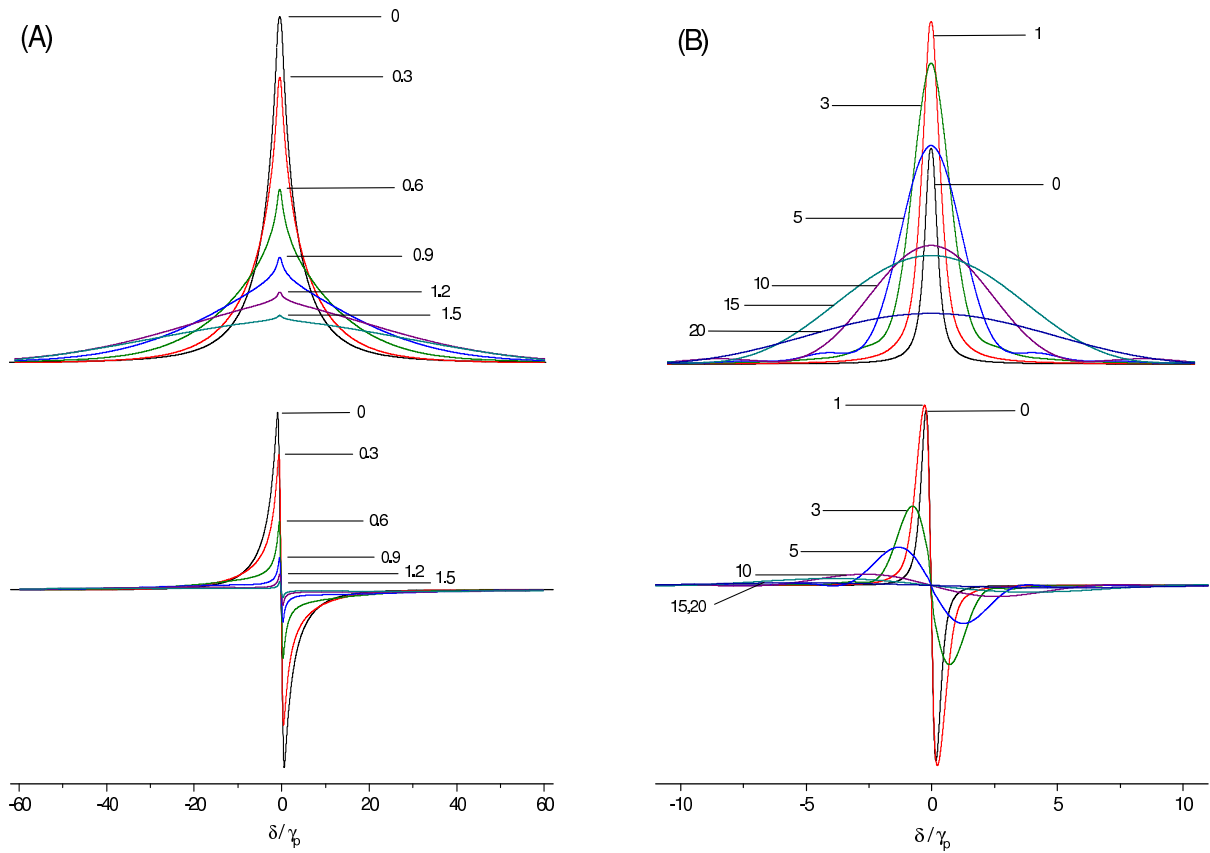}
\caption{\label{fig:detuning}(Color online) Dark resonance spectra
(top) and their derivative (bottom) for A) $\phi=0.1$ and B)
$\phi=10$ corresponding to different values of the laser detuning
$\frac{\delta\omega}{\Gamma}$ from the optical resonance.}
\end{figure*}
\subsection{\label{sbsec:level6} Role of the effective relaxation parameter $\gamma$.}
As was previously mentioned, in order to account for the finite beam
cross section as well as other possible sources of ground state
decoherence, a phenomenological decay rate $\gamma$ and a repumping
process $\lambda$ can be included in our model [see
Fig.~\ref{fig:levels} and Eqs.~(\ref{eq_1})]. In consequence, the
dark resonance spectrum is also characterized by the dimensionless
coherence loss rate $\frac{\gamma}{\gamma_p}$ (see Appendix).\\
Considering that $\frac{\gamma}{\gamma_p} < 1$, the parameter
$\gamma$ introduces a lower limit in the width of the dark resonance
spectrum for an open system as it is shown by the dashed line in
Fig.~\ref{fig:width}. The build-up of the narrow resonance can be
operated only during a mean time $\gamma^{-1}$, and as a consequence
the limiting value for $\Delta_{pp}$ appearing for long enough cells
depends on both rates $\gamma$ and $\gamma_p$. Such a limiting value
tends to $\Delta_{pp} = \gamma$ as the light intensity approaches
zero.\\
When $\frac{\gamma}{\gamma_p} \geq 1$ (i.e. strong decoherence
introduced by $\gamma$), the spectrum width is $\gamma$ dependent
for any cell length. In such a case, the mean atom-light interaction
time is essentially limited by $\gamma^{-1}$ and the velocity range
that gives an effective contribution to the spectra can be estimated
by $\gamma L$, as shown at Fig.~\ref{fig:velosel}.B(e).\\
\subsection{\label{sbsec:level7} Effect of the laser detuning.}
We here briefly address the situation of a nonzero detuning of the
optical field ($\delta\omega \neq 0$). Once again the two different
situations $\phi < 1$ and $\phi > 1$ should be distinguished as
illustrated in Fig.~\ref{fig:detuning}. In the first one, a narrow
contribution to the dark resonance spectrum is observable provided
that the slowest atoms (corresponding to the velocity range
$\phi\frac{\Gamma}{k}$ around zero) remains within the optically
selected velocity class ($\delta\omega\leqslant\Gamma$). As
$\delta\omega$ is increased the narrow contribution to the dark
resonance spectrum decreases as a consequence of detuning
(Fig.~\ref{fig:detuning}.A). Also, a broad structure appears in the
dark resonance spectrum for $\delta\omega \gtrsim \Gamma$. This
structure corresponds to optically selected fast atoms that have a
small interaction time and thus a broader contribution to the
spectra.\\
In the case $\phi > 1$, the range of velocities enabling an atom to
contribute to the dark resonance spectrum increases. If
$\frac{\delta\omega}{\Gamma} \lesssim \phi$ a fraction of this
velocity range is optically selected and a narrow contribution to
the spectrum is observed, as shown in Fig.~\ref{fig:detuning}.B
(note in this figure that the amplitude of the dark resonance
spectrum of the open system for $\frac{\delta\omega}{\Gamma}=0$ is
smaller than the one for $\frac{\delta\omega}{\Gamma}=1$ due to
population losses that are more efficient at resonance; such an
effect does not exist for a closed system).\\
\section{\label{sec:level8} Influence of the velocity distribution.}
Most of the results presented in the preceding sections crucially
depend on the assumption that the velocity distribution is a M-B
function. For a thin cell, the atomic mean free path in free space
exceeds the distance between walls, so that the bulk $v_z$
distribution should be strongly connected to the $v_z$ distribution
for the particles at desorption.\\
A M-B distribution is theoretically predicted for the velocities of
particles desorbed from a plane surface, when the surface and the
gas are in thermodynamical equilibrium \cite{Gaede:1913}. From the
flux of desorbing atoms, one expects to recover a M-B velocity
distribution, implying a $cos\theta$ law (lambertian law) of
probability for the velocity at a given angle $\theta$ with respect
to the perpendicular to the desorption surface. Interestingly, this
standard assumption has been addressed experimentally only seldomly.
Experimental data exist from an accurate linear absorption
experiment \cite{Grischkowsky:1980}, where a thermodynamical
equilibrium between atoms and surface was expected, and from
experiments using molecular beams scattered by planar surfaces
\cite{Bordo:1999}. In this last case, the distribution is usually
referred as Knudsen law, as no thermodynamical equilibrium exists
between the incoming particles and the surface.\\
\begin{figure*}[!] \centering
\includegraphics [width=16cm]{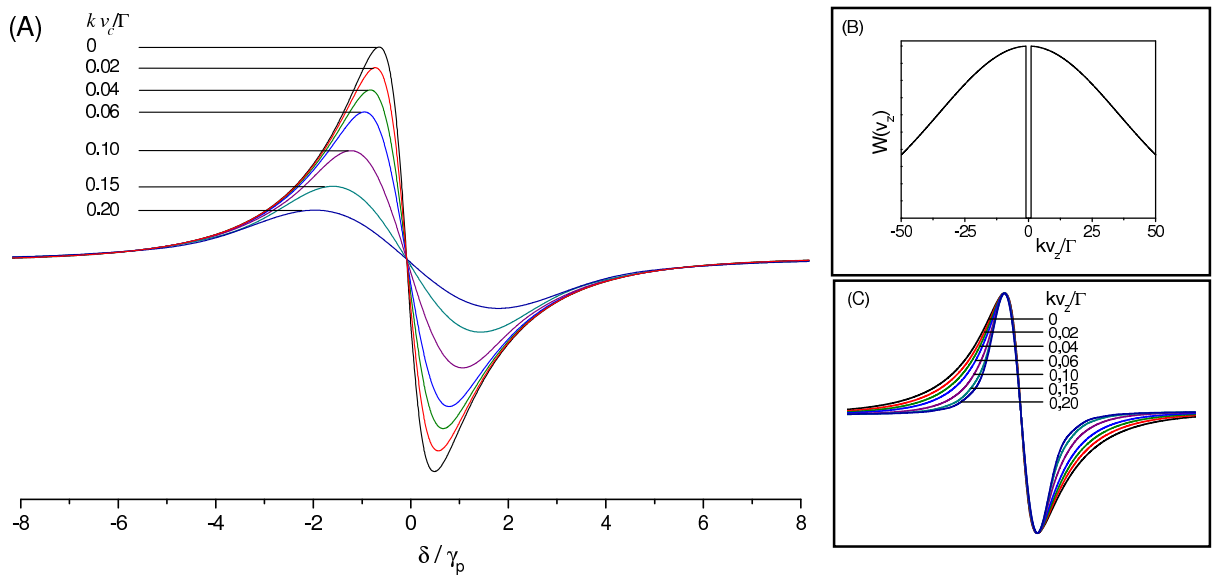}
\caption{\label{fig:dist}(Color online) A) Evolution of the
derivative of the dark resonance spectrum, peak-to-peak width and
amplitude when velocities $|v_z|$ smaller than $\frac{kv_c
}{\Gamma}$ are removed from the M-B velocity distribution; as
illustrated in B) for $\frac{kv_c }{\Gamma} = 1$. The spectra were
calculated for realistic parameters easily attainable in experiments
($kL=40, \Omega=0.1\Gamma (\phi=0.4), \gamma=0.001\Gamma$ and
$\alpha=0.7$). For the purpose of lineshapes comparison, in C) the
curves are shown normalized to have the same peak-to-peak amplitude
and width.}
\end{figure*}
According to Fig.~\ref{fig:velosel}.B the velocities contributing to
the dark resonance spectrum in micro-metric thin cells are much
smaller than the optically selected velocity $\frac{\Gamma}{k}$, and
are currently well below $0.1\frac{\Gamma}{k}$. For alkali atoms,
such as Rb or Cs, this corresponds to longitudinal velocities
smaller than $v_z=1\: m/s$. The existence of such slow atoms,
necessarily assumed when considering a M-B distribution, is
questionable. For example, even minor effects of surface roughness
or imperfections in the wall planarity (i.e. deviation between the
local normal direction, and the global one) may prohibit the
observation of very small velocities at desorption. In addition, at
the limit of very slow velocities, the quantized atomic motion, and
the Heisenberg uncertainty casts a limit to the minimal velocity if
the atomic position is approximately known. In all these cases, and
even if collisional redistribution may allow atoms with nearly null
$v_z$ velocity between the walls, the transient evolution of the
atomic state can no longer be described by a wall-to-wall
trajectory. Moreover, atoms leaving the surface at a nearly-grazing
incidence undergo a strong van der Waals interaction for a long
duration, susceptible to deviate their trajectories. Note that such
a possible limitation to the description by a M-B distribution,
affecting very slow atoms, could not be addressed in the experiments
mentioned above for these extremely small velocities. Similarly, in
an atomic scattering experiment, an $80^\circ$ angle relatively to
the normal (i.e. $v_z \simeq 0.2 \overline{v}$) is usually an
already very large scattering angle. The present results show the
importance of the contribution of slow atoms, both in amplitude and
for the lineshape. Hence one understands that dark resonances in
thin cells may provide a useful tool to investigate the presence of
atoms with ultra-slow normal velocities resulting from a desorption
process. To analyze this possibility we have studied the
modifications of the dark resonance spectrum when the slowest atoms
are removed from the velocity distribution. In such case, the narrow
structure of the coherent signal at $\delta= 0$ is correspondingly
reduced, the spectrum broadens and changes its shape as shown in
Fig.~\ref{fig:dist}. In the specific case shown in
Fig.~\ref{fig:dist}, obtained for a realistic set of parameters
($\phi = 0.4$, $\gamma=0.001\Gamma$ has been used corresponding to
experiments with Rb or Cs \cite{Figueroa:2006}), a significant
change in the dark resonance spectrum shape and width is already
visible when velocities up to $v_z = 0.1 \frac{\Gamma}{k}$ are
removed from the velocity distribution. Moreover, the changes
resulting from the modified velocity distribution are not limited to
a change in amplitude and width, that could be hardly identified if
one has no possibility to effectively compare the obtained signal
with the one predicted for a genuine MB distribution. Rather, it is
the overall shape of the signal, notably in the ratio between the
wings and the behavior at the center, that are modified. Conversely,
in purely optical spectroscopy, the contribution of atoms with
ultra-slow velocities, although relatively favored in thin cells
\cite{Briaudeau:1999, Briaudeau:th} and depending upon the same
$\phi$ parameter, remains hindered inside the (broad) optical width.
Here, the dark resonance scheme introduces extra parameters, finally
allowing an RF selectivity in an all-optical measurement. It is also
worth noticing that, although going to long cells is apparently
favorable to reduce the width, the sensitivity to low velocity atoms
is rather obtained for relatively short cells as shown by
Fig.~\ref{fig:dist}.

\section{\label{sec:level6}Conclusions}

In summary, we have studied the dark resonances in micro-meter thin
cells identifying the relevant physical parameters and discussing
the physical mechanisms that determine the shape of the spectra. We
have mostly assumed the irradiation to be on resonance with the
atomic transition, but we have shown this is not critical (within
the optical width). For large values of the parameter $\phi$
(corresponding to relatively long cells and/or high intensities)
narrow dark resonance spectra result from the contribution of all
atoms : in spite of the assumption of wall-to-wall trajectory, one
hence recovers the results predicted for a macroscopic cell.  For
small values of $\phi$ (corresponding to relatively thin
micro-metric cells and low intensities), broader dark resonance
spectra are obtained that, nonetheless, should enable an unusual
tight velocity selection of ultra-slow atoms.  This differs from
single laser beam experiments on a two-level transition, for which
the predicted enhanced contribution of slow atoms (and according to
the same parameter $\phi$) remains hindered by the optical width
\cite{Briaudeau:th}. In our case, it is the EIT scheme that allows
to benefit of an RF selectivity in an all-optical measurement. We
suggest that dark resonance spectroscopy in thin cells could provide
a useful tool for the investigation of the velocity distribution of
atoms leaving the surface with grazing trajectories after a
desorption process. This possibility, expected from spectroscopy in
thin cell \cite{Briaudeau:1996} (and demonstrated in
\cite{Briaudeau:2000} down to $2 m/s$, as compared to the $5 m/s$
standard selection associated to the optical width), was until now
limited by the optical width of the detection method
\cite{Briaudeau:1999}, but turns here to be realistic with the
introduction of the arbitrary narrow Raman width as an additional
parameter. Other interesting phenomena such as velocity or
temperature jumps on a surface \cite{Goodman:1976} or laser induced
desorption \cite{Bordo:1999} could also possibly be approached with
this technique. The theoretical analysis presented in this work
shows a slight dependence on the dark resonance spectral width with
the confinement introduced by the thin cell that can notably apply
to the consideration of thin cells for the development of compact
atomic frequency references \cite{Failache:2007}. Finally,
experiments related to this work are currently underway on Rb or Cs,
in which inhomegeneities and/or finite transit time impose a
practical lower limit for the dark resonance spectral width. However
a spectral width of only $\Delta_{pp} \simeq 200 kHz$ have been
measured for the hyperfine Raman transition of Rb in a $10 \mu m$
thin cell
\cite{Failache:2007}.\\

\section{\label{sec:level7}Acknowledgements}
We would like to thank financial support from ECOS-Sud (contract
$\#U00-E03$, France) and CSIC (Uruguay).\\

\section{Appendix: Parameters characterizing the dark resonance spectra in a thin cell}
In this appendix we present the analytic procedure followed to
identify the set of parameters ($\frac{\Omega^2kL}{\Gamma^2}$,
$\frac{\delta}{\gamma_p}$, $\frac{\gamma}{\gamma_p}$ and $\alpha$)
that characterizes the dark resonance spectra in thin cells.\\
Defining the real 9-elements vector
\begin{equation}
\begin{split}
\bm{\sigma}&
=(\sigma_{DD},\sigma_{CC},\sigma_{ee},Re\sigma_{eD},Im\sigma_{eD},Re\sigma_{eC} \\
& ,Im\sigma_{eC},Re\sigma_{DC},Im\sigma_{DC},)
\end{split}
\end{equation}
the Bloch equations (Eq.~\ref{eq_1}) can be written in the following
matrix form
\begin{equation}
\dot{\bm{\sigma}}=\bm{M}\bm{\sigma}+\bm\lambda   \label{eq_2}
\end{equation}
where $\bm\lambda=(\frac{\gamma}{2}\:\frac{\gamma}{2}\:0\:0\:0
\:0\:0\:0\:0)$. After solving the eigen-values and eigen-vectors
problem we have: $\bm{M} =\bm{V}\bm{D}\bm{V}^{-1}$ where $\bm{D}$ is
the diagonal matrix of eigen-values $\epsilon_i$ and $\bm{V}$ is a
transformation matrix (that in particular is not dependent on $z$
due to the symmetry of the system of Fig.~\ref{fig:levels}.A), the
spatial integral in Eq.~\ref{eq_5} can be written
\begin{equation}
\begin{split}
& \bm{m}.\bm{V}\left(\int_{0}^{kL}\exp\left(\frac{\bm{D}kz}{kv_z}
\right)\;dkz\right)\bm{V}^{-1}\bm\sigma_o\; \\
& +\bm{m}.\bm{V}\left(\int_{0}^{kL}\left(\exp\left(\frac{\bm{D}kz
}{kv_z}\right)-\bm{I}\right)\;dkz\right)\bm{D}^{-1}\bm{V}%
^{-1}\bm\lambda \label{eq_6}
\end{split}
\end{equation}
where $\bm m=(000000100)$ allows the notation $Im\sigma_{eC}=\bm
m.\bm\sigma $. After calculating the spatial integral in
Eq.~\ref{eq_6}, Eq.~\ref{eq_5} can be written
\begin{widetext}
\begin{equation}
\begin{split}
\label{eq_7} \Delta I \propto \Omega\;kL \int_0^{+\infty}
W(v_z)\;dv_z\;\bm m .\bm V
\left[\left(\exp\left(\frac{\bm{D}kL}{kv_z}\right)-\bm{I}\right)(\bm{D}kL)^{-1}
\bm{V}^{-1}\bm\sigma_o\;\delta_D\right]+\\
+\Omega^3{kL}^2 \int_0^{+\infty} W(v_z)\;dv_z\;\bm m .\bm V
\left[\left(\exp\left(\frac{\bm{D}kL}
{kv_z}\right)-\bm{I}\right)(\bm{D}kL)^{-1}kv_z-\bm{I}\right](\bm{D}kL)^{-1}
\bm{V}^{-1}\frac{\bm\lambda}{\Omega^2}
\end{split}
\end{equation}
\end{widetext}
\begin{figure}[!]
\includegraphics[width=8.6cm]{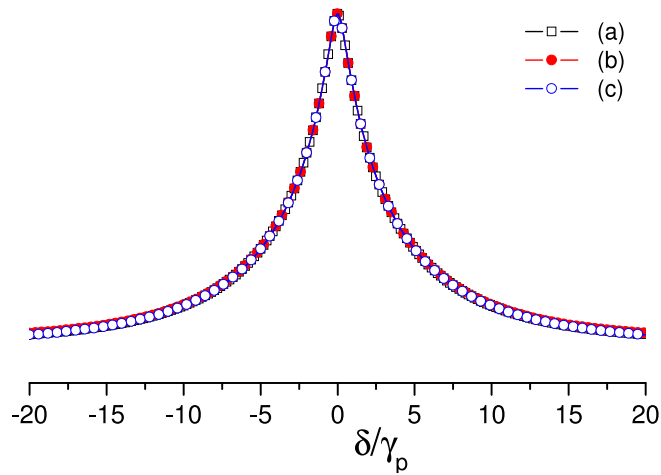}
\caption{\label{fig:compara}(Color online) Numerically calculated
dark resonance spectra from Eq.~(\ref{eq_7}) for $\phi=0.1$,
$\frac{\gamma}{\gamma_p}=0.01$ and $\alpha=0.7$ corresponding to (a)
$kL=1000$, $\Omega=0.01 \Gamma$, $\gamma=1\times10^{-6}\Gamma$ (b)
$kL=250$, $\Omega=0.02 \Gamma$, $\gamma=4\times10^{-6}\Gamma$ and
(c) $kL=25000$, $\Omega=0.002 \Gamma$, $\gamma=4\times10^{-8}
\Gamma$. The same vertical scale is used to show all spectra.}
\end{figure}
We have solved the problem of finding the analytic expression of
most of the eigen-values of $\bm M$ by a perturbation procedure on
$\Omega$ and $\delta$, considering then $\delta \ll \Gamma$ and
$\Omega \ll \Gamma$. The matrix $\bm M$ of Eq.~(\ref{eq_2}) have
eigen-values around $\frac{-\Gamma}{2}$ and $-\Gamma$ associated to
a fast optical evolution at rates $\frac{\Gamma}{2}$ and $\Gamma$
and eigen-values around zero, associated to a slow evolution related
to the Raman coherence evolution. The dark resonance spectra are
essentially determined by the rates of these last eigen-values that
have a general expression, after the perturbative calculation,
\begin{equation}
\epsilon _i\approx \emph{a}_i+\emph{b}_i\;\Omega^2+\emph{c}_i
\left(\frac{\delta}{\Omega}\right) ^2  \label{eq_8}
\end{equation}
for $\delta \ll \Omega$. The constant $\emph{a}_i$ is equal to
$-\gamma $ or $\frac{-\gamma}{2}$ and the constants $\emph{b}_i$ and
$\emph{c}_i$ depend only on the parameter $\alpha$. Introducing
~(\ref{eq_8}) in the expression $\bm{D}kL$ we found for each term of
the diagonal of $\bm D$ the generic form,
\begin{equation}
\epsilon _i\;kL=\left(\frac{\Omega^2
kL}{\Gamma^2}\right)\left(\frac{\emph{a}_i}{\gamma_p}+
\emph{b}_i\;+\emph{c}_i\left(\frac{
\delta}{\gamma_p}\right)^2\right) \label{eq_9}
\end{equation}
After ~(\ref{eq_9}) the exponents in ~(\ref{eq_7}) depend on the
parameters $\frac{\gamma_p kL}{\Gamma}$, $\frac{\delta}{\gamma_p}$,
$ \frac{\gamma}{\gamma_p}$ and $\alpha$, suggesting a dimensionless
set of parameters that could determine the dark resonance spectra.
We have verified by solving numerically the Eq.~\ref{eq_7} that
these parameters effectively characterize the dark resonance spectra
[see Fig.~\ref{fig:compara}].

\end{document}